\documentstyle[aps,prl,twocolumn,floats,epsfig]{revtex}

\begin{document}
\twocolumn[\hsize\textwidth\columnwidth\hsize\csname @twocolumnfalse\endcsname
\title{Micro-Hall Magnetometry Studies of Thermally Assisted and Pure Quantum 
Tunneling in Single Molecule Magnet Mn$_{12}$-Acetate}
\author{Louisa Bokacheva \cite{louisa} and Andrew D. Kent \cite{andy}}

\address{Department of Physics, New York University,
4 Washington Place, New York, New York 10003}

\author{Marc A. Walters \cite{marc}}

\address{Department of Chemistry, New York University,
31 Washington Place, New York, New York 10003}

\date{December 12, 2000}

\maketitle

\begin{abstract}
We have studied the crossover between thermally assisted and pure quantum 
tunneling in single crystals of high spin ($S=10$) uniaxial single molecule magnet 
Mn$_{12}$-acetate 
using micro-Hall effect magnetometry. Magnetic hysteresis experiments have been used to 
investigate the energy levels that determine the magnetization reversal as a function 
of magnetic field and temperature. These experiments demonstrate that the crossover 
occurs in a narrow ($\sim 0.1$~K) or broad ($\sim1$~K) temperature interval depending on 
the magnitude and direction of the applied field. For low external fields applied 
parallel 
to the easy axis, the energy levels that dominate the tunneling shift abruptly with 
temperature. In the presence of a transverse field and/or large longitudinal field 
these energy levels change with temperature more gradually. A comparison of our 
experimental results with model calculations of this crossover suggest that there are 
additional mechanisms that enhance the tunneling rate of low lying energy levels and 
broaden the crossover for small transverse fields.
\end{abstract}
\vskip0.5cm
\centerline{{\small Keywords: Single-Molecule Magnets; High-Spin Clusters; Quantum 
Tunneling; Hysteresis; Hall-Effect Magnetometry}}
\vskip0.5cm
]


High spin single molecule magnets Mn$_{12}$-acetate (Mn$_{12}$) and Fe$_8$ have been 
actively studied as model systems for the behavior of the mesoscopic spins 
\cite{Chudnovsky,Friedman,Ohm,Sessoli,Novak,Hernandez,Wernsdorfer,EPR1,EPR2,EPR3,NS,CCNS}. 
These materials can be considered as monodisperse ensembles of weakly interacting particles 
with net spin $S=10$ and strong uniaxial anisotropy. They enable studies of both 
classical and quantum effects through macroscopic magnetic measurements. These 
clusters show enhanced relaxation of magnetization at regular intervals of magnetic 
field, attributed to the tunneling of magnetization across the anisotropy barrier 
\cite{Friedman,Ohm}. 

The temperature dependence of this process suggests that both classical thermal 
activation and quantum tunneling are important \cite {Novak}. Other significant recent 
experimental results on single molecule magnets include observation of non-exponential 
relaxation of magnetization \cite{Ohm} and quantum phase interference in Fe$_8$ 
\cite {Wernsdorfer}. 
Spectroscopic experiments (EPR and inelastic neutron scattering) have provided 
important information about the magnetic energy levels of Mn$_{12}$ and Fe$_8$, relevant 
to understanding their macroscopic magnetic response \cite {EPR1,EPR2,EPR3,NS,CCNS}.

\begin{figure}[tb]
\vskip0.3cm
\centerline{\epsfig{file=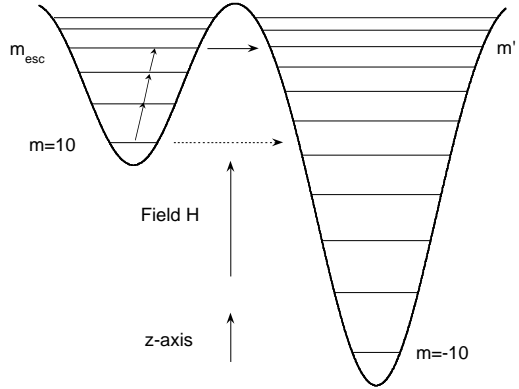,width=7cm}
}
\vskip0.3cm
\caption{Schematic of the double well potential in Mn$_{12}$. Solid arrows show 
magnetization reversal by thermally assisted tunneling, with thermal activation to an 
excited level in a metastable well and tunneling between resonant levels. The dashed 
arrow illustrates pure quantum tunneling.}
\label{fig1}
\end{figure}

Experimental results, such as the series of regular steps in magnetic hysteresis 
curves of these materials, have been interpreted within an effective spin Hamiltonian 
for an individual cluster:
\begin{equation}
{\cal H}=-DS_z^2 - BS_z^4 - g_z\mu_{\rm B} S_zH_z +{\cal H}',
\label{Ham}
\end{equation}
where the uniaxial anisotropy parameters $D$ and $B$ have been determined by EPR 
\cite{EPR3} and 
inelastic neutron spectroscopy experiments \cite{NS} ($D=0.548(3)$~K, $B=1.17(2)\times
10^{-3}$~K, and 
$g_z$ is estimated to be $1.94(1)$). Here $\cal H'$ includes terms which do not commute 
with $S_z$ 
and produce tunneling. These mechanisms of level-mixing may be due to a transverse 
field (such as hyperfine, dipolar fields or an external field, contributing terms like 
$H_xS_x$) or higher order transverse anisotropies, for example 
$C(S_+^4+S_-^4)$, 
$C=2.2(4) \times 10^{-5}$~K \cite{NS}, 
which is the lowest order term allowed by the tetragonal symmetry of the Mn$_{12}$ 
crystal. The enhanced relaxation of the magnetization at certain field values is 
ascribed to thermally assisted tunneling (TAT) or pure quantum tunneling (QT) (see 
Fig. \ref{fig1}). In the TAT regime the relaxation occurs by tunneling from 
thermally excited 
magnetic sublevels ($m_{\rm esc}=9, 8, ... ,  -8, -9$), when two levels on the opposite 
sides 
of the anisotropy barrier are brought close to resonance by the magnetic field. From 
the unperturbed Hamiltonian (1) the longitudinal ($z$-axis) field at which the levels 
$m_{\rm esc}$ and $m'$ become degenerate is:
\begin{equation}
H(n, m_{\rm esc})=nH_0 \{1+B/D [ m_{\rm esc}^2+(m_{\rm esc}-n)^2 ]\},
\label{StepPos}
\end{equation}
where  $n=m_{\rm esc}+m'$ is the step index describing the bias field, and 
$H_0=D/g_z\mu_{\rm B}$ is a constant ($0.42$~T). Direct numerical diagonalization of 
the spin Hamiltonian (1) shows 
that the small transverse anisotropy term does not significantly change these resonance 
fields. 
Note that a larger magnetic field is necessary to bring lower lying sublevels into 
resonance. As the temperature decreases, the thermal population of the excited levels 
is reduced, and these states contribute less and less to the tunneling. Consequently, 
the steps in hysteresis curves shift to higher bias field values, and steps with 
larger $n$ become observable. At low temperature, the tunneling from the ground state 
$m_{\rm esc}=10$ dominates, and the position and amplitude of the steps become 
independent of 
temperature, denoted the QT regime.

At any temperature the magnetic relaxation rate is dominated by the 
contribution of only few (one or two) magnetic sublevels. This is because the 
relaxation rate is the product of the thermal occupation probability and the tunneling 
rate. The probability of thermal occupation decreases exponentially with energy, while 
the tunneling probability increases exponentially approaching one at the top of the 
anisotropy barrier. Therefore the relaxation rate has a sharp maximum at the dominant 
level or levels.

Recent theoretical models suggest that in the large spin limit a crossover between 
thermal activation and quantum tunneling can either occur abruptly, in a narrow 
temperature interval, or gradually, in a broader temperature interval 
\cite{Crossover,Garanin,Crosstheory}. 
In the first case the energy at which the system crosses the anisotropy barrier shifts 
abruptly with temperature from a value close to the 
top of the barrier to the lowest lying level in the metastable well 
(denoted a first-order crossover). 
In the second case this energy changes 
smoothly with temperature (second-order) \cite{Note}. 
The ``phase diagram'' for this crossover 
depends on the form of the spin Hamiltonian, particularly the terms important for 
tunneling. In finite spin systems the crossover is always smeared, nevertheless the 
two types of the crossovers can be distinguished experimentally. In the first-order 
crossover there are competing maxima in the relaxation rate versus energy and the global 
maximum shifts abruptly 
from one energy to the other as a function of temperature. In the second-order case 
a single maximum in the relaxation rate shifts continuously with temperature. Recent 
experiments have shown that in Mn$_{12}$ the crossover occurs in a narrow temperature 
interval when the applied field is parallel to the easy axis of the sample 
\cite{OurEPL}. 
In contrast, experiments on Fe$_8$ suggest a second-order crossover 
\cite{WernsdorferEPL}.

In this paper we show that in Mn$_{12}$ both types of the crossover can be 
experimentally 
observed when the applied field has a transverse component. A transverse magnetic 
field makes the crossover more gradual and leads to a continuous shift in the dominant 
energy levels (i.e., a second-order crossover) \cite{PRL}. We show that these levels can 
be identified by hysteresis experiments and their behavior as a function of temperature 
can be studied in order to test the tunneling model described above.

Our experiments have been conducted using a micro-Hall-effect magnetometer \cite{HallBar}
in a high field $^3$He  system. Single crystals of Mn$_{12}$ in the shape of 
parallelepipeds $50 \times 50 \times 200$ $\mu$m$^3$  
were synthesized as described in Ref.~\cite{Lis}. A crystal was 
encapsulated in thermally conducting grease and the temperature was measured with a 
calibrated carbon thermometer a few millimeters from the sample. 
The angle $\theta$ between 
the easy axis of the crystal and the applied magnetic field was varied by rotating the 
sample in a superconducting solenoid.

\begin{figure}[t]
\centerline{\epsfig{file=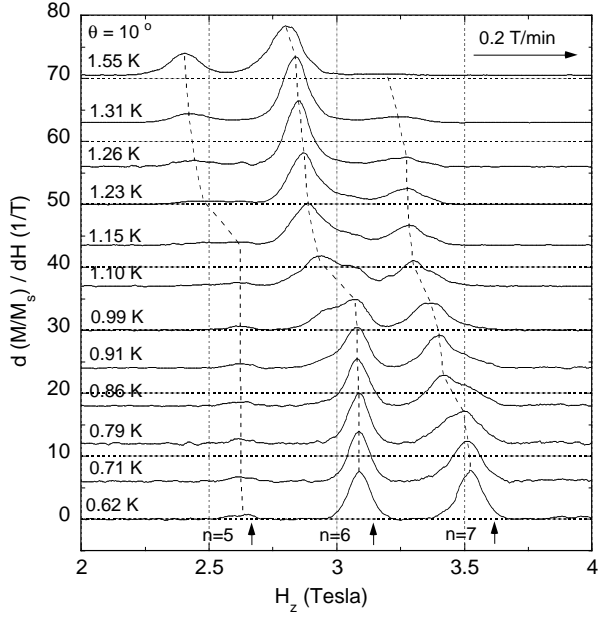,width=0.450\textwidth}
}
\vskip0.3cm
\caption{Field derivative of normalized magnetization $d(M/M_s)/dH$ vs $H_z$ at different 
temperatures measured at $\theta=10^\circ$ between the applied field and the easy 
axis of the 
crystal. The curves are offset for clarity. The dashed lines mark the positions of the 
maxima in $dM/dH$. Arrows correspond to the resonance fields $H(n, m_{\rm esc}=10)$ for 
$n=5$, $6$ 
and $7$ calculated according to Eq. (2).}
\label{fig2}
\end{figure}

In our hysteresis experiments the sample was initially saturated ($M_0=-M_{\rm s}$), 
then 
the field was ramped at a constant rate ($0.2$~ T/min) towards positive saturation. 
Hysteresis curves show steps and plateaus, separated by a field interval of 
approximately $0.44$ T, in agreement with previously published results. Figure 
\ref{fig2} shows a 
plot of the derivative of the magnetization $dM/dH$ versus the applied field for 
$\theta=10^\circ$
measured at different temperatures. 

The positions and structure of the peaks in $dM/dH$ 
show the magnetic fields at which there are maxima in the magnetization relaxation 
rate at a given temperature, applied field and magnetization. The dashed lines mark 
the positions of the three observed peaks showing their shift with temperature. The 
peak that occurs at the lowest magnetic field shifts gradually from $H=2.41$~T to 
$H=2.48$~T (by $0.17H_0$) as the temperature decreases from $1.55$~K to $1.23$~K, 
that can be identified as the TAT regime. 

As the temperature is reduced from $1.23$~K to $1.15$~K, 
this maximum is shifted abruptly to $H=2.62$~T (i.e., by $0.34H_0$) and remains 
approximately at the same
value upon further decrease of temperature to $0.62$~K. The middle peak also shifts 
gradually from $2.81$~T to $2.94$~T in the interval between $1.55$~K to $1.10$~K, 
but it then shifts abruptly between $1.10$~K and $0.91$~K, at lower temperature than 
the previous 
maximum. Below $0.91$~K the position of this peak also remains constant. We attribute 
this temperature independent regime to the pure QT. Assuming that at the lowest 
temperature only the ground state $m_{\rm esc}=10$ participates in tunneling, we can 
identify 
the positions of the observed peaks by the resonance field values 
$H(n, m_{\rm esc}=10)$, $n=5$, 
$6$, and $7$, calculated according to Eq. (2), from low to high magnetic field. 
The peaks 
labeled $n=5$ and $6$ show an abrupt crossover between TAT to QT, which occurs in an 
interval of approximately $0.1$ K.
In contrast to $n=5$ and $n=6$ peaks, the $n=7$ peak shifts to higher field 
(from $3.20$~T to 
$3.52$~T) step-wise in the whole studied temperature interval ($\sim 1$~K). 
We consider the crossover to QT in this case as gradual (second-order).

\begin{figure}[t]
\vskip0.3cm
\centerline{\epsfig{file=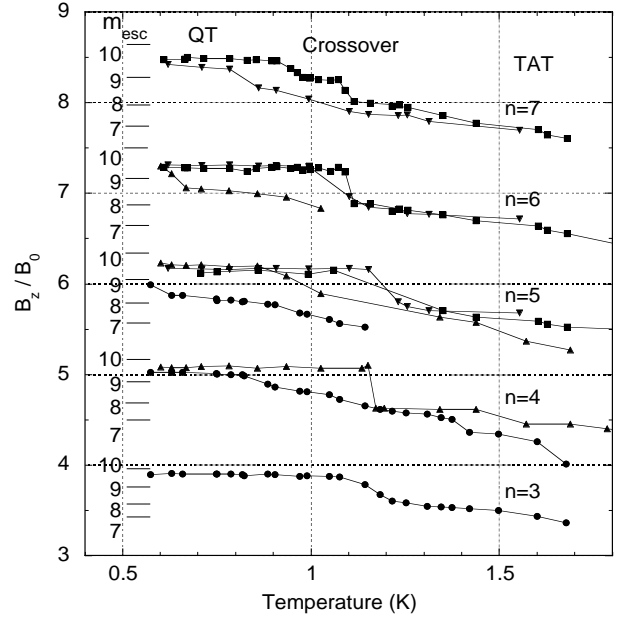,width=0.45\textwidth,height=0.48\textwidth}
}
\vskip0.3cm
\caption{Peak positions (in the units of $B_0=0.42$ T) versus temperature for 
$\theta=0^\circ$ 
(squares), $\theta=10^\circ$ (triangles down), $\theta=20^\circ$ (triangles up), 
$\theta=35^\circ$ (circles). 
The bars on 
the left hand side of the graph show the escape levels calculated using Eq. (2). The 
accuracy with which the peak positions can be determined is approximately the size of 
the symbol.}
\vskip0.3cm
\label{fig3}
\end{figure}

Peak positions as a function of temperature, obtained from hysteresis experiments 
performed as described above, are summarized in Fig. \ref{fig3}. 
This plot shows the values of 
the longitudinal field, at which the maxima in $dM/dH$ occur, versus 
temperature for the 
four studied angles, $\theta=0^\circ$, $10^\circ$, $20^\circ$, and $35^\circ$.
The peak positions were corrected for the 
effects of the internal field as described in Ref. \cite{PRL}. 
The bars on the left hand 
side of the figure show the escape levels calculated using 
Eq. (2), with parameters 
from spectroscopic data \cite{EPR3,NS}.

Analyzing this graph, we can make following observations. First, for larger angles, 
and therefore higher transverse field, peaks with lower indices (i.e., lower $H_z$) can 
be observed in the experimental time window.  The lowest step observed for  
$\theta=0^\circ$ and 
$10^\circ$ is $n=5$, for $\theta=20^\circ$ it is $n=4$, for $\theta=35^\circ$ it is 
$n=3$. 
This is consistent with the idea 
that the transverse field promotes tunneling and lowers the effective anisotropy 
barrier. We find that there is greater amplitude in lower lying peaks as the 
transverse field is increased. Second, two regimes can be distinguished: the high 
temperature regime, where the peaks gradually shift to higher fields with decreasing 
temperature, and the low temperature regime, where the peak positions are constant. We 
associate the first regime with the TAT and the second with pure QT. Third, the form 
of the crossover between these two regimes depends on the longitudinal field. For each 
sample orientation, peaks with lower indices show a more abrupt crossover between TAT 
and QT than peaks with higher indices (compare peaks $n=6$ and $n=7$ for $\theta=0^\circ$
and  $10^\circ$, or 
$n=4$ and $5$ for $\theta=20^\circ$, or $n=3$ and $4$ for $\theta=20^\circ$).

\begin{figure}[t]
\centerline{
\epsfig{file=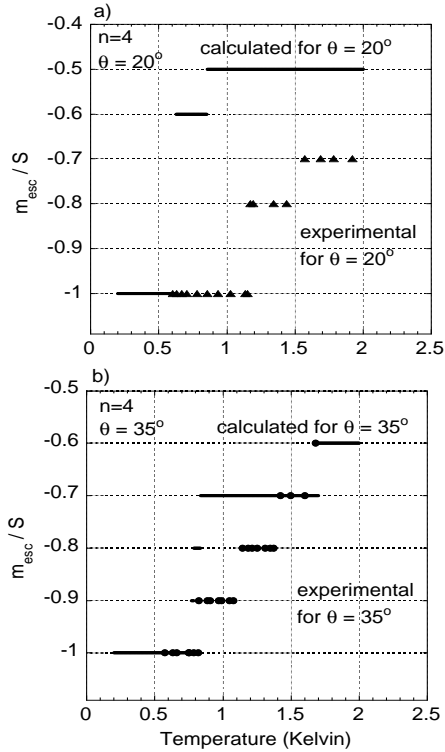,width=6cm,height=10cm}
}
\vskip0.5cm
\caption{Experimental escape levels $m_{\rm esc}$ as a function of temperature 
for $n=4$ at two 
different orientations: 
a)~$\theta=20^\circ$ (triangles), and b) $\theta=35^\circ$ (circles). 
Solid lines show 
$m_{esc}$ calculated by numerical diagonalization of the Hamiltonian (1) 
with $D=0.66$ K, 
$B=0$ K.}
\label{fig4}
\end{figure}

The experimentally determined escape levels can be compared to models 
\cite{Crossover,Garanin,Garanin1}. As 
mentioned above, the relaxation rate can be written as a product of the tunneling rate 
between the pair of levels $i$ and $j$, split by the energy $\Delta_{ij}$, and the 
Boltzmann factor: \linebreak
$\Gamma_{ij} \sim \Delta_{ij}^2 \exp({-E_{ij}/T}$), where $E_{ij} = (|E_i + E_j|)/2 - E_0$  
is the average energy of levels $i$ 
and $j$ above the lowest lying state in the metastable well $E_0$. The energy level 
$m_{\rm esc}$ 
which dominates tunneling is determined as the level for which the tunneling rate $\Gamma$ 
has a maximum at a given temperature, longitudinal and transverse field. The tunnel 
splittings $\Delta_{ij}$ were obtained by numerical diagonalization of spin 
Hamiltonian (1), 
whose eigenvalues were determined with high precision \cite{Garanin1}. 
In Eq. (1) the second 
order uniaxial anisotropy term (i.e., $BS_z^4$) was ignored for simplicity and a higher 
value of $D$ ($D=0.66$~K) was used so that the low lying energy level spacing and overall 
barrier height is better approximated. Figure \ref{fig4} shows experimental positions of the 
$n=4$ peak at two different angles, $\theta=20^\circ$ (Fig. 4(a)) and $\theta=35^\circ$
(Fig.~4(b)) compared to 
the calculated $m_{\rm esc}$. Calculations were performed using a Mathematica program for 
constant values of the longitudinal and transverse magnetic field corresponding to 
$n=4$, $\theta=20^\circ$ and $\theta=35^\circ$.

First consider the $\theta=20^\circ$ data. A few major discrepancies between 
the calculated and 
experimental results can be noticed. The experimental crossover temperature is higher 
than the calculated one ($T_{\rm exp}=1.25$~K versus $T_{\rm th}=0.7$~K). 
Also the 
experimental 
temperature interval in which the crossover occurs is larger than predicted by the 
model. Experimental escape levels involved in the crossover ($m_{\rm esc}=7-10$) are lying 
lower in the potential well than the calculated ones ($m_{\rm esc}=5-10$). The observed 
escape 
levels change more gradually than predicted by the model: our observations show that 
none of the levels $m_{\rm esc}=7-9$ are skipped in the crossover, while the calculations 
show 
that as many as three levels ($m_{\rm esc}=7, 8, 9$) do not contribute significantly to the 
relaxation. Based on this comparison we conclude that there are additional mechanisms, 
such as transverse anisotropy, that enhance the tunneling rate of low lying levels.
For $\theta=35^\circ$ the crossover is more gradual than for $\theta=20^\circ$, in 
qualitative 
agreement with 
the theoretical model. As in the $\theta=20^\circ$ case, experimental data show that 
lower lying 
levels ($8$ and $9$) are active over larger intervals of temperature than 
predicted by the model. The overall better agreement with the calculated data 
in this case suggests that a large enough applied transverse field can become 
a dominant factor in determining the tunnel splittings.

\begin{figure}[t]
\centerline{
\epsfig{file=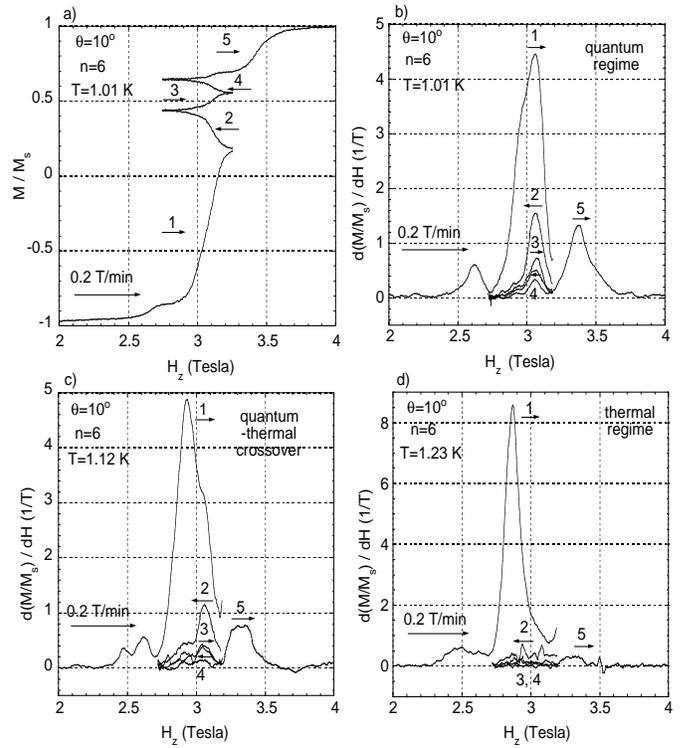,width=0.5\textwidth,height=10cm}
}
\vskip0.5cm
\caption{Minor loop hysteresis experiments for $\theta=10^\circ$, $n=6$: 
a) typical hysteresis 
curve at $T=1.01$ K; b), c), d) normalized $d(M/M_s)/dH$ at $T=1.01$ K, $1.12$ K, 
and $1.23$~K 
respectively. The direction and order of the field sweeps is indicated by numbered 
arrows.}
\label{fig5}
\end{figure}

It is important to note that in our hysteresis experiments, in which the field is 
increasing, relaxation will first occur from thermally excited states because these 
come into resonance first (Eq. (2)). Therefore our data on Figures \ref{fig3} and \ref{fig4}
somewhat underestimate the crossover temperature.

To show this we have performed a series 
of minor loop hysteresis experiments near the crossover temperature. In these 
experiments the field was swept at a constant rate back and forth across an interval, 
where a peak in $dM/dH$ was observed. If the field is ramped towards positive 
saturation, relaxation from the excited levels is emphasized, if the direction of the 
sweep is towards lower magnetization, the ground state tunneling is favored.
The results of these experiments in which the field was swept across the n=6 peak five 
times at $\theta=10^\circ$ are shown on Fig. \ref{fig5}. Figure 5(a) shows a typical hysteresis 
curve at 
$T=1.01$~K. Figures 5(b-d) show plots of $dM/dH$ at three different temperatures. For 
$T=1.01$~K (Fig. 5(b)) the maximum in $dM/dH$ for all sweeps occurs at $3.06$~T, which 
corresponds to QT from $m_{\rm esc}=10$. At a higher temperature, $T=1.12$~K (Fig. 5(c)) 
the 
maximum of the first sweep is at a lower field, $H=2.93$~T, which can be identified as a 
thermal channel ($m_{\rm esc}= 9$). For the sweeps $2$ - $5$ the maximum occurs at the QT 
position 
($3.06$ T), which suggests that at this temperature QT still dominates. At higher 
temperature ($1.23$ K, Fig. 5(d)) almost all relaxation occurs via a thermal channel at 
$H=2.87$~T ($m_{\rm esc}=8$), which shows that the crossover to TAT has already taken 
place. 
These results show that regular hysteresis experiments emphasize the tunneling from 
the excited states, and therefore the crossover temperature determined from such 
experiments is underestimated.

In conclusion, we presented low temperature magnetic hysteresis studies of thermally 
assisted and pure quantum tunneling in Mn$_{12}$. The crossover between these two regimes 
can be either abrupt or gradual. These types of the crossovers are distinguished by the 
temperature interval within which the escape energy levels 
shift from thermally excited levels to the lowest state in the metastable well. 
Through the hysteresis experiments we were able to identify these energy levels and 
study their behavior as a function of temperature. Our studies indicate that the 
higher longitudinal and transverse fields make the crossover more gradual. Comparison 
of our experimental data and theoretical model suggests that in Mn$_{12}$ there must be 
additional mechanisms which promote tunneling, such as a transverse anisotropy, which 
is responsible for broadening the crossover for smaller transverse fields and 
increasing the tunneling rate from low lying levels. These experiments establish a 
lower bound on the crossover temperature (Fig. 3), as the standard hysteresis 
experiments favor relaxation from the excited levels.

We are grateful to Dmitry Garanin for fruitful discussions and help with numerical 
calculations.



\vspace{-0.2cm}
\begin{references}
\vspace{-1.3cm}
\bibitem[*]{louisa}
louisa@mailaps.org

\bibitem[**]{andy}
andy.kent@nyu.edu

\bibitem[\dagger]{marc}
marc.walters@nyu.edu

\bibitem{Chudnovsky}
see, E. M. Chudnovsky and J. Tejada, {\it Macroscopic tunneling of the magnetic moment}  
(Cambridge University Press, Cambridge, UK 1997), Chapter 7.

\bibitem{Friedman}
J. R. Friedman, M. P. Sarachik, J. Tejada, and R. Ziolo, Phys. Rev. Lett. {\bf 76}, 
3830 (1996); L. Thomas, F. Lionti, R. Ballou, D. Gatteschi, R. Sessoli, and B. Barbara, 
Nature {\bf 383}, 145 (1996).

\bibitem{Ohm}
C. Sangregorio, T. Ohm, C. Paulsen, R. Sessoli, D. Gatteschi, Phys. Rev. Lett. 
{\bf 78}, 4645 (1997).

\bibitem{Sessoli}
R. Sessoli, D. Gatteschi, A. Caneschi, and M. A. Novak, Nature {\bf 365}, 141 (1993). 

\bibitem{Novak}
M. Novak and R. Sessoli, in {\it Quantum Tunneling of Magnetization-QTM'94}, 
ed. by L. Gunther and B. Barbara (Kluwer Publishing, Dordrecht, 1995) p. 171; 
B. Barbara, W. Wernsdorfer, L. C. Sampaio, J. G. Park, R. Ferr\'{e}, L. Thomas, C.Paulsen,
M. A. Novak, A. Benoit, K. Hasselbach, D. Mailly, R. Sessoli, and A. Cansechi, J. Magn.
Magn. Mater. {\bf 140-144}, 1825 (1995).

\bibitem{Hernandez}
J. M. Hernandez, X. X. Zhang, F. Luis, J. Bartolom\'{e}, J. Tejada, and R. Ziolo, 
Europhys. Lett. {\bf 35}, 301 (1996).

\bibitem{Wernsdorfer}
W. Wernsdorfer and R. Sessoli, Science {\bf 284}, 133 (1999).

\bibitem{EPR1}
S. Hill, J. A. A. J. Perenboom, N. S. Dalal, T. Hathaway, T. Stalcup, and J. S. Brooks, 
Phys. Rev. Lett. {\bf 80}, 2453 (1998). 

\bibitem{EPR2}
M. Hennion, L. Pardi, I. Mirebeau, E. Suard, R. Sessoli, and A. Caneschi, 
Phys. Rev. B {\bf 56}, 8819 (1997).

\bibitem{EPR3}
A. L. Barra, D. Gatteschi, and R. Sessoli, Phys. Rev. B {\bf 56}, 8192 (1997).

\bibitem{NS}
I. Mirebeau, M. Hennion, H. Casalta, H. Andres, H. U. G\"{u}del, A. V. Irodova, and A.
Caneschi, Phys. Rev. Lett. {\bf 83}, 628 (1999).

\bibitem{CCNS}
Y. Zhong, M. P. Sarachik, J. R. Friedman, R. A. Robinson, T. M. Kelley, H. Nakotte, 
A. C. Christianson, F. Trouw, S. M. J. Aubin, and D. N. Hendrickson, J. Appl. Phys. 
{\bf 85}, 5636 (1999).

\bibitem{Crossover}
E. M. Chudnovsky and D. A. Garanin, Phys. Rev. Lett. {\bf 79}, 4469 (1997).

\bibitem{Garanin}
D. A. Garanin, X. Hidalgo, and E. M. Chudnovsky, Phys. Rev. B 57, 13639 (1998).

\bibitem{Crosstheory}
G.-H. Kim, Phys. Rev. B {\bf 59}, 11847 (1999); G.-H. Kim, Europhys. Lett. {\bf 51}, 
216 (2000); H. J. W. M\"{u}ller-Kirsten, D. K. Park, and J. M. S. Rana, Phys. Rev. B 
{\bf 60}, 6662 (1999), and references therein.

\bibitem{Note}
The analogy to phase transitions is a purely formal one as discussed in 
\cite{Crossover}. The first-order, second-order terminology for the escape rate crossover 
is originally due to Larkin and Ovchinnikov \cite{Larkin}.

\bibitem{Larkin}
A. I. Larkin and Y. N. Ovchinnikov, Sov. Phys. JETP {\bf 59}, 420 (1984).

\bibitem{OurEPL}
A. D. Kent, Y. Zhong, L. Bokacheva, D. Ruiz, D. N. Hendrickson, and M. P. Sarachik, 
Europhys. Lett. {\bf 49}, 512 (2000).

\bibitem{WernsdorferEPL}
W. Wernsdorfer, R. Sessoli, A. Caneschi, D. Gatteschi, and A. Cornia, 
Europhys. Lett. {\bf 50}, 552 (2000).

\bibitem{PRL}
L. Bokacheva, A. D. Kent, and M. A. Walters, Phys. Rev. Lett. {\bf 85}, 4803 (2000).

\bibitem{HallBar}
A. D. Kent, S. von Molnar, S. Gider, and D. D. Awschalom, J. Appl. Phys. {\bf 76}, 
6656 (1994).

\bibitem{Lis}
T. Lis, Acta Cryst. B {\bf 36}, 2042 (1980).

\bibitem{Garanin1}
D. A. Garanin and E. M. Chudnovsky, Phys. Rev. B {\bf 63}, 024418 (2001).
\end{references}
\end{document}